\documentclass{elsart}
\usepackage[dvips]{graphicx}

\begin{document}
\begin{frontmatter}
  \title{Achievement of 35 MV/m  in the Superconducting Nine-Cell
  Cavities for TESLA} 
  \date{\today}
  \author[DESY]{L. Lilje\corauthref{cor}},
  \corauth[cor]{Corresponding author.}
  \ead{Lutz.Lilje@desy.de}
  \author[KEK]{E. Kako},
  \author[DESY]{D. Kostin},
  \author[DESY]{A. Matheisen},
  \author[DESY]{W.-D. M\"oller},
  \author[DESY]{D. Proch},
  \author[DESY]{D. Reschke},
  \author[KEK]{K. Saito},
  \author[UNIHH]{P. Schm\"user},
  \author[DESY]{S. Simrock},
  \author[Nomura]{T. Suzuki},
  \author[DESY]{K. Twarowski}

  \address[DESY]{DESY, Notkestrasse 85, D-22607 Hamburg, Germany}
  \address[KEK]{KEK, High Energy Accelerator Research Organization, 1-1, Oho, Tsukuba, Ibaraki, 305-0801, Japan}
  \address[UNIHH]{Universit\"at Hamburg, Notkestrasse 85, D-22607
  Hamburg, Germany} 
  \address[Nomura]{Nomura Plating Co., Ltd., 5,  Satsuki, Kanuma, Tochigi, 322-0014, Japan} 

\begin{abstract}
The Tera Electronvolt Superconducting Linear Accelerator TESLA is the
only linear electron-positron collider project based on superconductor
technology for particle acceleration.  In the first stage with 500 GeV
center-of-mass energy an accelerating field of 23.4 MV/m is needed in
the superconducting niobium cavities which are operated at a
temperature of 2 K and a quality factor $Q_0$ of $10^{10}$. This
performance has been reliably achieved in the cavities of the TESLA
Test Facility (TTF) accelerator. The upgrade of TESLA to 800 GeV
requires accelerating gradients of 35 MV/m.  Using an improved cavity
treatment by electrolytic polishing it has been possible to raise the
gradient to 35 - 43 MV/m in single cell resonators. Here we report on
the successful transfer of the electropolishing technique to
multi-cell cavities. Presently four nine-cell cavities have achieved
35 MV/m at $Q_0 \ge 5\times 10^{9}$, and a fifth cavity could be
excited to 39 MV/m. In two high-power tests it
could be verified that EP-cavities preserve their excellent
performance after welding into the helium cryostat and assembly of the
high-power coupler. One cavity has been operated for 1100~hours at the
TESLA-800 gradient of 35 MV/m and 57 hours at 36 MV/m without loss 
in performance.

\end{abstract}

\maketitle

\begin{keyword}
 Superconducting RF cavities \sep Niobium \sep Surface
 superconductivity  \sep
  Accelerating gradients \sep High-energy accelerators \PACS 74.25.Nf
  \sep 74.60.Ec \sep 81.65.Ps \sep 84.70.+P \sep 29.17.+W
\end{keyword}
\end{frontmatter}

\section{Introduction}
\label{sec:introduction}
Electron-positron colliders have played a central role in the
discovery of new quarks and leptons and the formulation and detailed
verification of the Standard Model of elementary particle physics.
Circular colliders beyond LEP are ruled out by the huge synchrotron
radiation losses, increasing with the fourth power of energy. Hence a
linear collider is the only viable approach to center-of-mass energies
in the TeV regime. Such a linear lepton collider would be
complementary to the Large Hadron Collider (LHC) and allow detailed
studies of the properties of the Higgs particle(s). In the baseline
design of the superconducting TESLA collider \cite{tesla_cavity_tdr}
the center-of-mass energy is 500 GeV (TESLA-500), well above the
threshold for the production of the Standard-Model Higgs particle. The
possibility for a later upgrade to 800 GeV (TESLA-800) is considered
an essential feature to increase the research potential of the
facility for the study of supersymmetry and physics beyond the
Standard Model.

The design gradient of the TESLA-500 niobium cavities (23.4 MV/m) was
chosen at a time when the typical accelerating fields in
superconducting cavities were in the order of 3--5 MV/m. The factor of
five increase presented an ambitious goal which, however, has been
reached owing to the concentrated R$\&$D efforts of the TESLA
collaboration and of other institutions.  A detailed description of
the nine-cell cavities for the TESLA Test Facility (TTF) linac can be
found in \cite{tesla_cavities}. In the most recent series of 24
industrially produced TTF cavities the average gradient was measured
to be $25 \pm 2.6~$MV/m at a quality factor $Q_0 = 10^{10}$. 
After the fabrication 
process these resonators were cleaned at DESY by chemical etching 
(``buffered chemical polishing BCP'', see below) and subjected to a 
$1400^\circ$C heat treatment to enhance the low-temperature thermal 
conductivity of the niobium.

After many years of intensive R$\&$D there exists now compelling
evidence that the BCP process limits the attainable field in
multi-cell niobium cavities to about 30 MV/m, significantly below the
physical limit of about 45 MV/m which is given by the condition that
the rf magnetic field has to stay below the critical field of the
superconductor. For the type II superconductor niobium the maximum
tolerable rf field appears to be close to the thermodynamic critical
field (190 mT at 2 Kelvin).
 
Since a number of years, an improved preparation technique of the
inner cavity surface by electrolytic polishing (or
``electropolishing'' for short) has opened the way to gradients of
35~-~43~MV/m in 1.3 GHz single-cell cavities \cite{kako_99,lilje_99}.
This development motivated a thorough R\&D program on the
electropolishing (EP) of single-cell test cavities. The results have
been published recently \cite{lilje_03}.  In the present paper we
report on the successful transfer of the EP technology to the
nine-cell TESLA cavities.

\section{Preparation of the inner cavity surface}

\subsection{Chemical etching and electrolytic polishing}
\label{sec:etching}
Here we give a short outline of the chemical and electro-chemical
methods which are applied to clean and prepare the inner cavity
surface after the fabrication process. More details are found in
\cite{lilje_03}.  Niobium metal has a natural Nb$_2$O$_5$ layer with a
thickness of about 5~nm which is chemically rather inert and can be
dissolved only with hydrofluoric acid (HF).  The sheet rolling of
niobium produces a damage layer of about 100 $\mu$m thickness that has
to be removed in order to obtain a surface with excellent
superconducting properties. One possibility is chemical etching which
consists of two alternating processes: dissolution of the Nb$_2$O$_5$
layer by HF and re-oxidation of the niobium by a strongly oxidizing
acid such as nitric acid (HNO$_3$) \cite{gmelin_nb,siemens_scrf}.  To
reduce the etching speed a buffer substance is added, for example
phosphoric acid H$_3$PO$_4$ 
\cite{guerin_bcp2}, and the mixture is cooled below 15$^\circ$C.  The
standard procedure with a removal rate of about 1~$\mu$m per minute is
called {\it buffered chemical polishing} (BCP) with an acid mixture
containing 1~part~HF (40\%), 1~part~HNO$_3$ (65\%) and
2~parts~H$_3$PO$_4$ (85\%) in
volume.  At TTF, a closed-circuit chemistry system is used in which
the acid is pumped from a storage tank through a cooling system and a
filter into the cavity and then back to the storage.

A gentler preparation method is provided by electrolytic polishing
(EP). The material is removed in an acid mixture under the flow of an
electric current. Sharp edges are smoothed out and a very glossy
surface can be obtained.  The electric field is high at protrusions so
these will be dissolved readily while the field is low in the
boundaries between grains and little material will be removed
here. This is an essential difference to the BCP process which tends
to enhance the steps at grain boundaries.

The electro-chemical processes are   as follows
\cite{kneisel_kfk,ponto}:
\begin{eqnarray}
  2 \textrm{Nb} + 5 \textrm{SO}_4^{--} + 5 \textrm{H}_2\textrm{O}& \to &
  \textrm{Nb}_2\textrm{O}_5 + 10 \textrm{H}^+ + 5\textrm{SO}_4^{--}
  + 10 \textrm{e}^- \nonumber \\
  \textrm{Nb}_2\textrm{O}_5 +6 \textrm{HF} & \to &
  \textrm{H}_2\textrm{NbOF}_5 + \textrm{NbOF}_2 \cdot 0.5
  \textrm{H}_2\textrm{O} + 1.5 \textrm{H}_2\textrm{O} \nonumber  \\
  \textrm{NbOF}_2 \cdot 0.5 \textrm{H}_2\textrm{O} + 4 \textrm{HF} &
  \to & \textrm{H}_2\textrm{NbF}_5 + 1.5 \textrm{H}_2\textrm{O} \nonumber
\end{eqnarray}

The  roughness of electropolished niobium surfaces 
is less than $0.1~\mu$m \cite{claire_99} 
while chemically etched surfaces are
 at least an order of magnitude rougher. The main 
 advantage of EP is  the far better 
smoothening of the ridges at grain boundaries.  An
electropolishing of at least 100~$\mu$m is needed both
for surface smoothening and damage layer removal.

\subsection{Electropolishing of nine-cell cavities}
\label{sec:cavity_ep}
At the KEK laboratory a long-term experience exists with the
electropolishing of multi-cell cavities. The five-cell 508~MHz
cavities of the TRISTAN electron-positron storage ring
\cite{saito_ep_system} were electropolished by an industrial company
(Nomura Plating).  Within a joint KEK-DESY program 9 TESLA nine-cell
resonators of the most recent industrial production have been
electropolished at Nomura Plating.  The EP parameters are summarized in
table \ref{tab:ep_parameters}.
\begin{figure}
\begin{center}
  \includegraphics[angle=90,width=10cm]{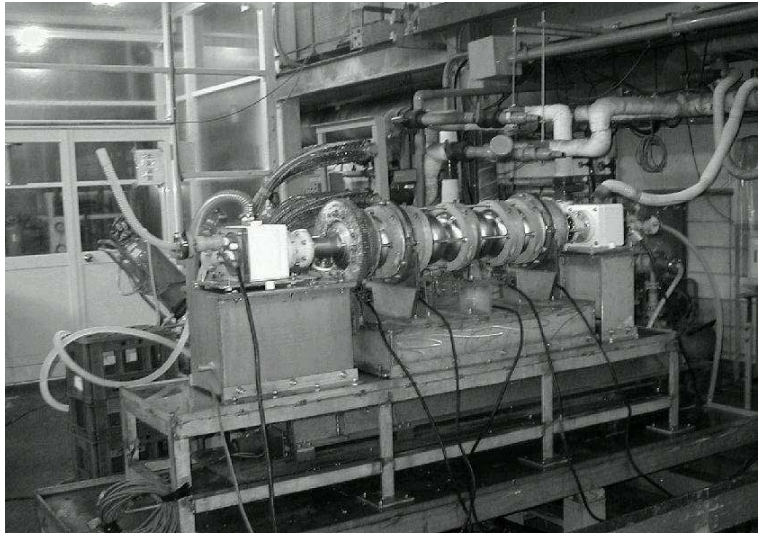}  
\caption{Setup for the electropolishing of multi-cell cavities at Nomura
  Plating (Japan).}
\label{fig:ep_scheme} 
\end{center}
\end{figure}
\begin{table}[btp]
  \centering
  \begin{tabular}[t]{|l|l|}
\hline
Acid mixture      & 10 \% ~ HF (40\%) \\
               & 90 \% ~ H$_2$SO$_4$ (96\%) \\
\hline
Voltage                    & 15 -20      ~ V \\
Current density            & 0.5 -0.6      ~ A/cm$^2$ \\
Removal rate               & 30         ~ $\mu$m/hour\\
Temperature of electrolyte & 30 - 35     ~ $^\circ$C \\
Rotation                   & 1           ~ rpm \\
Acid flow                  & 5           ~ liters/s \\
\hline
  \end{tabular}
  \caption{Parameters for the EP of nine-cell cavities.}
  \label{tab:ep_parameters}
\end{table}

The cavity is installed horizontally (see figure
\ref{fig:ep_scheme}) together with the aluminum cathode. The lower half of the
cavity is filled with the electrolyte which reacts with the the
niobium surface only very slowly  when no voltage is applied (etch
rate less than 1 nm per hour).  
After the equilibrium filling level has been reached, 
the cavity is put into  rotation and the  voltage is applied between
 cavity and cathode while the current-voltage relationship is
monitored. 
At a voltage of 15 - 20 V,  the current through the
electrolyte starts to oscillate indicating that the following two processes
are taking place in alternating order: dissolution of the
Nb$_2$O$_5$ by HF and re-oxidation of the Nb by H$_2$SO$_4$. 
 The best polishing
results are obtained for a current oscillation of 10~-~15\% about the
mean value.  The temperature of the acid mixture is kept in the range
30~-~35$^\circ$C. Temperatures   above $40^\circ$C must be
avoided as they result in etching pits on the surface.  When the
desired amount of material has been removed, the current is switched
off. The rotation is stopped and the cavity is put into vertical
position to drain the acid mixture. After rinsing with pure water the
electrode is dismounted while keeping the cavity filled with water,
thus avoiding drying stains from acid residues. The cavity is then
transported into a clean room for high-pressure water rinsing.
 
An electropolishing facility for nine-cell cavities has recently been
commissioned at DESY. The electrolyte is circulated in a closed
loop. The cathode is made from pure aluminum and is surrounded with a
tube made from a porous PTFE\footnote{Polytetrafluoroethylene, for
example Teflon\textregistered} cloth to prevent the electrolytically
produced hydrogen from reaching the niobium surface.  Except for the
cathode all components of the EP system are made from chemically inert
plastic materials, e.g.  Polyperfluoro Alkoxyethylene (PFA),
Polyvinylidene Fluoride (PVDF) or PTFE.  The acid mixture is stored in
a Teflon-cladded container and water cooled via a Teflon-covered heat
exchanger. The electrolyte is pumped with a membrane pump through a
cooler and a filter with 1 $\mu$m pore size into the hollow cathode
which has openings at the centers of the cavity cells.  Inside the
cavity the volume above the electrolyte is filled with dry nitrogen to
prevent water vapour absorption by the strongly hygroscopic
H$_2$SO$_4$. The exhaust gases are pumped through a neutralization
system to avoid environmental hazards.

\section{Performance measurements on electropolished nine-cell cavities} 
\label{sec:ep_measurements}
\subsection{Low-power tests in the accelerating mode}
All nine electropolished cavities were cleaned at DESY
by rinsing with ultrapure water at high
pressure. After the drying  in a class-100 clean room, a low-power
input coupler  and a pickup antenna were mounted and the
cavity was closed with UHV vacuum flanges.  A first performance test
was carried out in a vertical bath cryostat filled with superfluid
helium at 2 Kelvin.  In this cryostat, the rf power of a few 100 watts
is transmitted into the cavity through a  movable antenna in the beam 
pipe section of the cavity. The external quality factor $Q_{ext}$ is
adjusted to be close to the intrinsic quality factor $Q_0 \approx 10^{10}$.
The time constant of the cavity is determined by the ``loaded quality factor''
$$ \tau=\frac{Q_L}{\omega_0}
\quad {\rm with} \quad Q_L=(\frac{1}{Q_0}+\frac{1}{Q_{ext}})^{-1} \approx \frac{Q_0}{2} \;.$$
A typical value is $\tau \approx 1~$s. 
When the cavity is operated in the pulsed mode the intrinsic quality
factor can be easily computed from the time decay of the stored
energy. The coupling strength  of the pickup antenna to the electric field
inside the cavity is determined by  pulsed measurements at 
low gradient. Once  this calibration  is known the excitation curve
$Q_0(E_{acc})$ can be measured in the  continuous wave (cw) mode. 

In the low-power test, two cavities showed strong field emission at
15-17 MV/m. These were sorted out for a second EP which is planned to
take place in the recently commissioned EP facility at DESY (see below).

The remaining seven cavities without strong field emission, having
passed the low-power test successfully, were then  evacuated to a
pressure of $10^{-7}$ mbar and subjected to a 48-hour bake-out at
$120^\circ$C. According to the experience with the single-cell cavities
\cite{lilje_99,lilje_03,visentin_bake_98,kneisel_99} this bakeout
is an essential prerequiste for achieving 
high gradients in electropolished cavities. After the bakeout the
performance tests in the vertical bath cryostat were repeated. The
results of these second tests are discussed in the following. 

\begin{figure}[t!]
\begin{center}
  \includegraphics[angle=0,width=8cm]{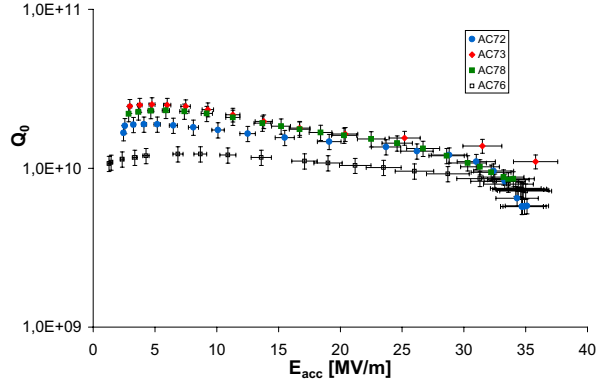}  
\caption[Excitation curves of electropolished nine-cell cavities
  with 'in-situ' bakeout.]{Excitation curves of the 4 best 
  electropolished nine-cell cavities after the EP at Nomura
  Plating. Plotted is the quality factor  
  $Q_0$ as a function of the accelerating field. The tests
  have been performed at 2~K.}
\label{fig:ep_ninecell_baked} 
\end{center}
\end{figure}

\begin{figure}[h!]
  \centering
  \includegraphics[angle=-90,width=8.5cm]{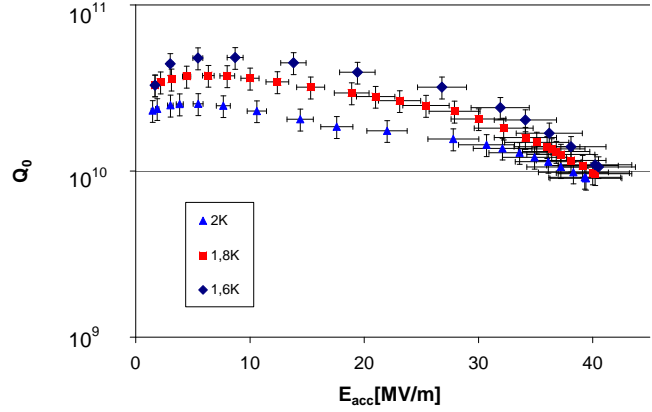}  
  \caption{Performance of a cavity which received a second EP at DESY. This
   is one of the cavities suffering from field emission after the  EP at
  Nomura Plating. An second EP of 40 $\mu$m removed the strong field
  emitter. A record gradient  of 39~MV/m at 2~K and of 40~MV/m below 1.8~K was
  achieved.}
  \label{fig:DESY_EP}
\end{figure}

\begin{figure}[h!]
  \centering
  \includegraphics[angle=-90,width=8.5cm]{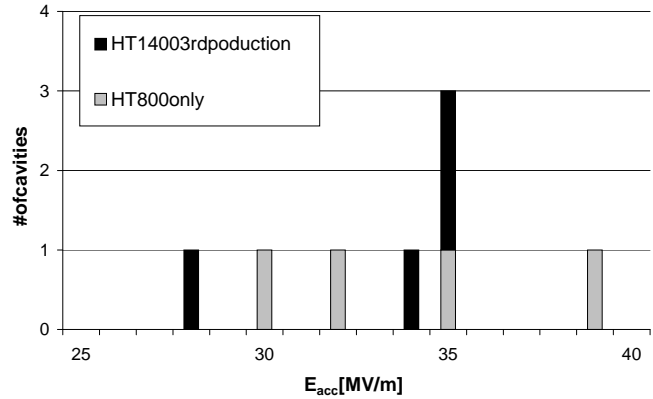}  
  \caption{Maximum accelerating field achieved in  electropolished
  nine-cell cavities. Gray bars: cavities with $800^\circ$C annealing,
  black bars: cavities with $800^\circ$C and $1400^\circ$C annealings
  in sequence. This is an indication that
  the annealing at $1400^\circ$C can be avoided for EP cavities, see sect. \ref{annealing}.}
  \label{fig:limitation_EP_ninecells}
\end{figure}

The excitation curves of the four best cavities after EP at KEK are
shown in figure \ref{fig:ep_ninecell_baked}. In November 2003 one of
the field-emission loaded cavities has been repolished 
 in the new EP facility at DESY. The test results of this
cavity at helium temperatures between 1.6 and 2.0 K are shown in
Fig. \ref{fig:DESY_EP} . Accelerating fields of up to 40 MV/m have
been reached which is a record for multicell niobium cavities. The
maximum accelerating fields achieved in all eight TTF cavities after
bakeout are shown in figure \ref{fig:limitation_EP_ninecells}.  
These results prove that the TESLA-800 gradient of  35 MV/m is  indeed
within reach.   

\begin{figure}[h!]
  \centering
  \includegraphics[angle=-90,width=8.5cm]{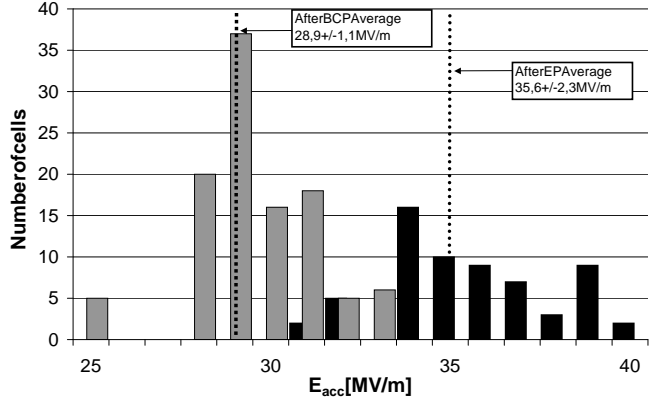}  
  \caption{Distribution of accelerating gradients in individual cells.
   The single cell statistics derived from the coupled  mode
  measurements are compared for  chemically etched (gray) and
  electropolished (black) nine-cell cavities. The average maximum
  gradient  is 28.9 MV/m for BCP-treated cavities and  35.6~MV/m for
  EP-treated cavities.} 
  \label{fig:compare_cells_EP_BCP}
\end{figure}

\pagebreak

\subsection{Performance of single cells in the nine-cell cavities}
In an $N$-cell cavity, each single-cell eigenmode splits up into $N$
coupled modes which are characterized by an rf phase advance of
$m\cdot \pi/N$ between neighbouring cells ($m=1, \ldots N$). For a
perfectly tuned cavity the normalized amplitudes ($A_{m,j}$) in the
individual cells are
\begin{equation}
A_{m,j}=\sqrt{\frac{2-\delta_{mN}}{N}}\,
\sin \left(\frac{m \pi}{2N}\,(2j-1)\right)
\end{equation}
where $m$ is the mode index, $j$ the cell number and $\delta_{mN}$ is
the Kronecker symbol ($\delta_{mN}=1$ for $m=N$ and 0
otherwise). Only in the $\pi$ 
mode with $m=N$ (the accelerating mode) the electric field has the
same magnitude in each cell. By measuring the excitation curves for
all coupled modes it is possible to determine the maximum attainable
field in each cell, apart from a left-right ambiguity: in a nine-cell
structure the cells 1 and 9, 2 and 8, 3 and 7, 4 and 6 are
indistuinguishable in this analysis.  The mode analysis is therefore a
useful tool to identify cells of lower performance and to enhance the
statistical basis for comparing the relative benefits of various
chemical or electro-chemical treatments. It should be noted, though,
that the electric field amplitudes in the cells depend critically on
slight frequency detunings from cell to cell. Therefore the
accelerating gradients derived from the mode analysis have a larger
systematic uncertainty ($\approx 15~\%$) than the gradient in the
accelerating $\pi$ mode which is determined with an accuracy of 8\%.

The single-cell performance of etched and electropolished cavities is
compared in figure \ref{fig:compare_cells_EP_BCP}.  In the BCP-treated
nine-cell TTF cavities of the third production series, the average
maximum gradient amounts to 28.9~MV/m. The electrolytically polished
cavities, on the other hand, achieve an average single-cell gradient
of 35.6~MV/m. (As mentioned above, one cavity with heavy
field-emission loading has been left out). This is clear evidence
that electropolishing is far superior to chemical etching in the
preparation of the rf surface of the cavities.

\begin{figure}[h!]
  \centering
  \includegraphics[angle=0,width=7cm]{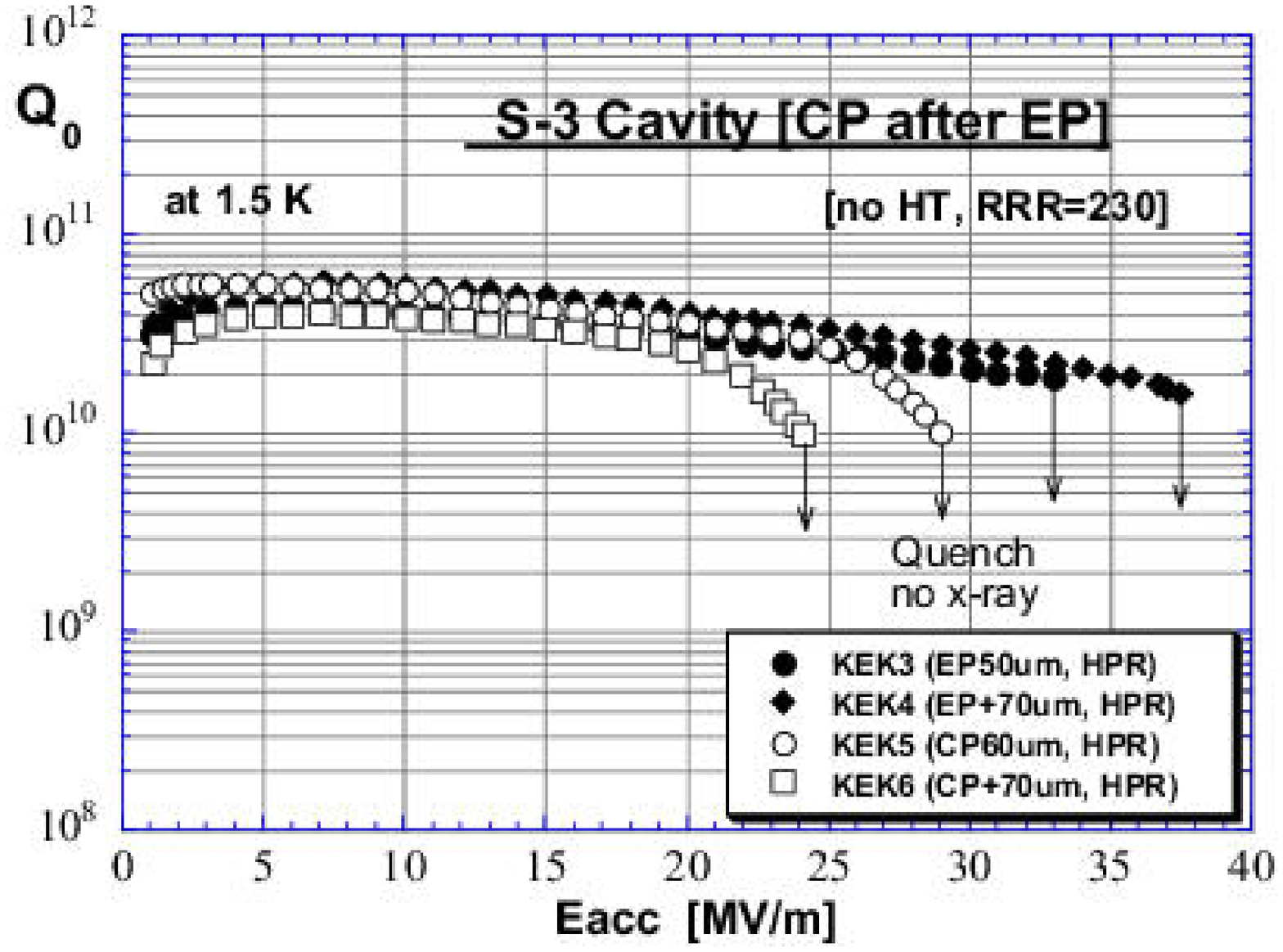}  \newline 
        a)

  \includegraphics[angle=0,width=6.6cm]{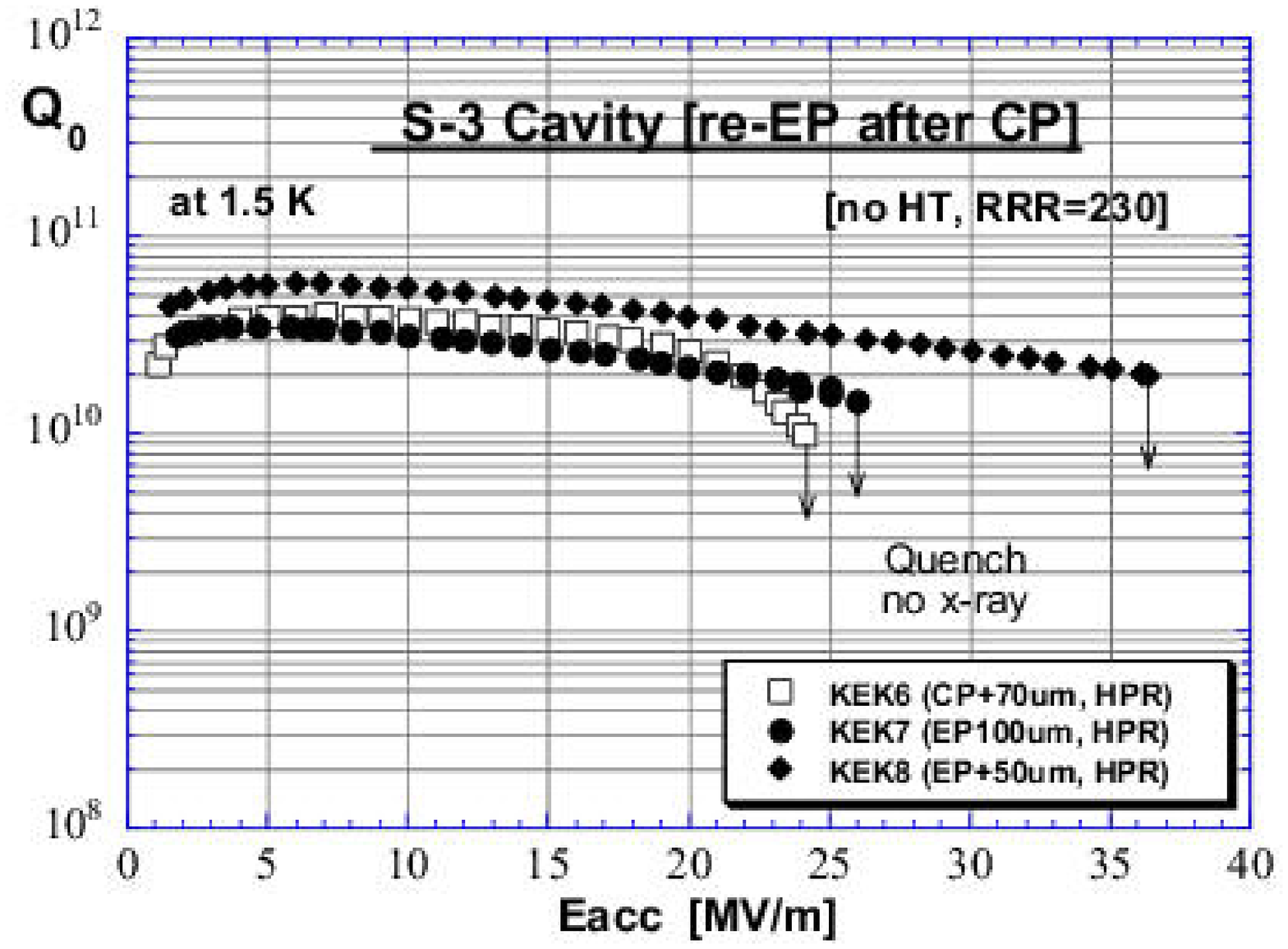}  \newline 
        b)
  \caption{Test series on a single-cell niobium
        cavity (S-3) at KEK: (a) Excitation curve
 of the cavity after EP (50 resp. 120 $\mu$m) 
and degradation due to chemical polishing (CP) after the EP. 
(b) Recovery of high gradient performance  due to a $150~\mu$m EP of
 the etched cavity. HPR stands for High Pressure
        Rinsing with ultrapure water.} 
  \label{fig:ep_after_bcp}
\end{figure}

\begin{figure}[h!]
  \centering
  \includegraphics[angle=-90,width=8cm]{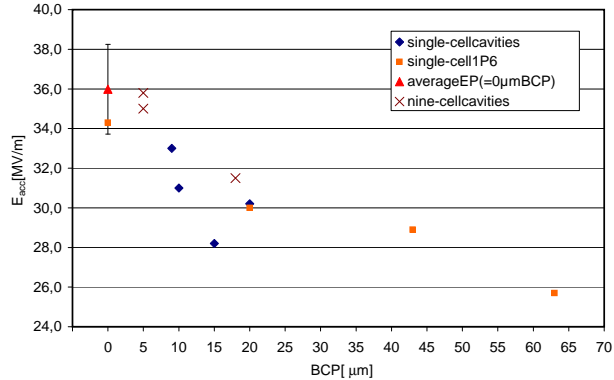}  
  \caption{Performance degradation of electropolished niobium cavities
  due to subsequent etching with a removal of 5 to 65 $\mu$m.}
  
  \label{fig:bcp_after_ep}
\end{figure}
The superiority of electropolishing was convincingly demonstrated in
an earlier experiment at KEK, see figure \ref{fig:ep_after_bcp}
\cite{kako_99}.  A single-cell (S-3) cavity reached 38 MV/m after an
EP $120~\mu$m. A subsequent chemical etching of 60 resp. 130 $\mu$m
reduced the gradient to 29 resp. 24 MV/m. 
A new electropolishing ($150~\mu$m) recovered the initial high
performance.  Studies at DESY on electropolished
single-cell and multi-cell cavities confirmed the performance
degradation due to a subsequent etching, see figure
\ref{fig:bcp_after_ep}.  An etching of just 20 $\mu$m 
 reduced the maximum accelerating gradient by 5 MV/m.

\subsection{Long-term stability of the electropolished surface}
\label{sec:bake_stable}
\begin{figure}[htbp]
  \centering
  \includegraphics[angle=0,width=10cm]{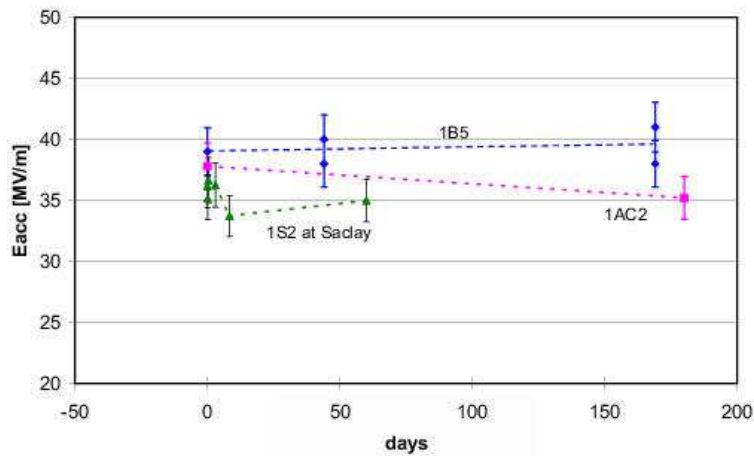}  
  \caption{Accelerating
    gradient as a function of exposure time to clean air. Within the
    measurement errors no difference in the behaviour of the cavities
    was observed.}
    \label{fig:air_exposure_new}
\end{figure}

\begin{figure}[htbp]
  \centering
  \includegraphics[angle=0,width=9cm]{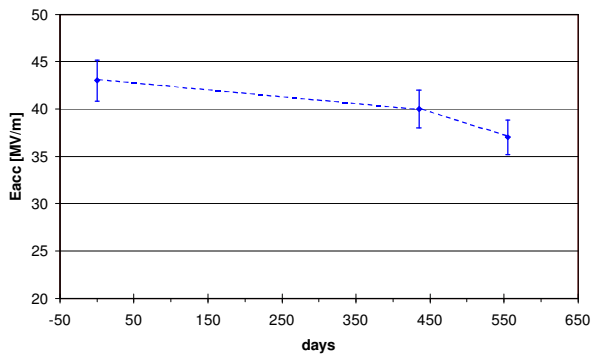}  
  \caption{Accelerating gradient as a function of exposure time to
  clean nitrogen. The observed reduction is still within the experimental
  errors. A second experiment is under way. Test temperature was 1.8~K.}
  \label{fig:nitrogen_exposure}
\end{figure}

The remarkable improvement of niobium cavities gained by
electropolishing will of course  only be useful for the accelerator if
the EP surface preserves its good properties over a long time period.
In a first endurance test at Saclay an EP-treated 1-cell cavity was
exposed to clean air for 2 months without a significant change in
performance.  This was verified in experiments at DESY (see figure
\ref{fig:air_exposure_new}) for a period of 6 months. One single-cell
cavity was filled with nitrogen and kept for more than 18 months. In
this case a reduction of the initial, very high gradient was observed
but the usable gradient was still above 35 MV/m, see 
Fig. \ref{fig:nitrogen_exposure}.

\subsection{High temperature heat treatments}
\subsubsection{Heat treatments of chemically etched cavities}
The TTF cavities are made from niobium sheets by deep drawing and
electron-beam welding. Two high-temperature heat treatments are
applied.  The first step is a two-hour annealing at 800$^{\circ}$C in
an Ultra High Vacuum (UHV) furnace which serves two purposes: removal of
dissolved hydrogen from the bulk niobium and mechanical stress
release from the deep drawn and welded multicell structure.  The
second step is a four-hour treatment at very high temperature
($1400^\circ$C) such that oxygen, nitrogen and carbon can diffuse out
of the niobium. A titanium layer is evaporated onto the niobium
surface to getter the foreign atoms and to protect the niobium
against oxidation by the residual gas in the UHV furnace. The
residual resistivity ratio $RRR$ and the low-temperature heat conductivity
of the bulk niobium increase by a factor of two, and the average
gradient in BCP-treated cavities grows by about 5 MV/m. The
experience at TTF has shown that the $1400^\circ$C heat treatment is
an indispensible prerequiste for achieving gradients above 25 MV/m
with a surface preparation by chemical etching (BCP).
 
\begin{figure}[htbp]
\begin{center}
  \includegraphics[angle=0,width=10cm]{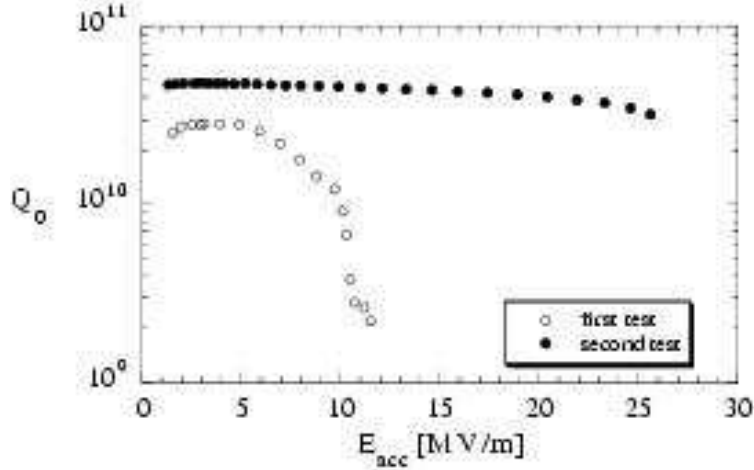} 
\caption{\label{Ti-removal}  Degradation of cavity performance 
by remainders of the titanium getter layer and
  improvement after complete   removal of this layer.}
\end{center}
\end{figure}

It must be noted, however, that the $1400^\circ$C annealing is
accompanied with a number of undesirable effects: large grain growth,
softening of the material, necessity of an additional etching to
remove the titanium getter layer. To obtain good deep drawing
properties the  grain size of the niobium sheets is in the order of
 50 $\mu$m. There is little grain growth at $800^\circ$C but
during the $1400^\circ$C annealing the grain size grows up to several
millimeters. The resulting tensile strength is very low, about 5 MPa,
hence the cavities are quite vulnerable to plastic deformation and
frequency detuning after the high-temperature treatment. Another
problem are the boundaries between the large grains. The titanium
getter penetrates deep into these boundaries, a depth of 60 $\mu$m has
been measured in samples. The titanium must be completely removed
since otherwise a drastically reduced performance is observed, see
Fig. \ref{Ti-removal}. For this reason the TTF cavities are subjected
to an 80 $\mu$m BCP after the $1400^\circ$C annealing.

\subsubsection{Heat treatments of electropolished cavities} 
\label{annealing}
The moderate annealing at around $800^\circ$C is an important cavity
preparation step which should be retained to avoid the danger of
niobium hydride formation (''Q-disease''), see \cite{lilje_03}.  On
the other hand, the experience gained at KEK and DESY with the
single-cell program justifies the expectation that the $1400^\circ$C
annealing can be safely discarded for electropolished cavities. 
The Japanese 1-cell cavities \cite{saito_99} were heat treated at
$T=760^\circ$C only, and in spite of a comparatively low residual
resistivity ratio of $RRR \le 300$, achieved gradients of 35 to 40
MV/m. In the CEA-CERN-DESY single-cell program \cite{lilje_03} seven
cavities were 
tested after the $800^\circ$C annealing with an average gradient of
35.4$\pm$5.3~MV/m and  three cavities were heat-treated at
$1400^\circ$C yielding 34.7$\pm$2.5~MV/m (see table \ref{tab:ep_ht}). 

The nine multicell TTF cavities sent to Japan for electrolytic
polishing had previously undergone the standard BCP treatment at TTF.
Five cavities had been annealed at $800^\circ$C while four cavities
were had been subjected to an additional $1400^\circ$C annealing.  The
field-emission loaded cavity which was sorted out belonged to
the first category.  The maximum gradients achieved in the eight
remaining 9-cell resonators have already been shown in
Fig. \ref{fig:limitation_EP_ninecells}, the average value is
$<E_{acc}>=34.0 \pm 3.9 ~$MV/m for the $800^\circ$C-annealed cavities
and $<E_{acc}>=33.0 \pm 3.3 ~$MV/m for the $1400^\circ$C-annealed
cavities. Two $800^\circ$C-annealed cavities belongs to the five
excellent nine-cell cavities shown in figures \ref{fig:ep_ninecell_baked}
and  \ref{fig:DESY_EP}.  From the single-cell analysis of the
eight cavities the following numbers are obtained: $<E_{acc}>=35.6 \pm
2.8 ~$MV/m for $800^\circ$C-annealing and $<E_{acc}>=35.6 \pm 1.7
~$MV/m for $1400^\circ$C-annealing (see table \ref{tab:ep_ht}).  These
results, combined with with 
the data from the single-cell R$\&$D programs, provide convincing
evidence that the $800^\circ$C annealing alone is sufficient to
achieve excellent performance in electropolished nine-cell cavities.

\begin{table}[btp]
  \centering
  \begin{tabular}[t]{|p{3.3cm}|c|c|}
\hline
Batch of cavities   & $800^\circ$C &$1400^\circ$C  \\
\hline
\hline
single-cells        & 35.4$\pm$5.3 & 34.7$\pm$2.5  \\
\hline
nine-cells          & 34.0$\pm$3.9 & 33.0$\pm$3.3  \\
\hline
single cell analysis of nine-cell cavities  & 35.6$\pm$2.8 & 35.6$\pm$1.7  \\
\hline

  \end{tabular}
  \caption{Overview on accelerating gradients measured on 
    electropolished cavities with different high temperature
    annealings.}  
  \label{tab:ep_ht}
\end{table}

In early 2004 a new series of nine-cell cavities will be delivered to
DESY. It is foreseen to prepare these cavities by electropolishing and
test them after the $800^\circ$C annealing. If the same promising
results should be obtained than found here, the $1400^\circ$C
annealing can certainly be avoided for the cavities of the proposed
X ray Free Electron Laser at DESY since this machine requires only
moderate acceleration fields.  Further investigations will be needed
to decide whether the $1400^\circ$C annealing would provide a larger
safety margin for the 800 GeV option of the TESLA
collider.

\subsection{High-power pulsed operation of electropolished cavities}
\subsubsection{Excitation curves}

In the TESLA collider the cavities have to be operated in the pulsed
mode to keep the heat load on the superfluid helium system within
acceptable limits. The rf power of about 210 kW per nine-cell
cavity (for TESLA-500) is transmitted through a coaxial power
coupler. The external quality factor amounts to $Q_{ext}=2.5\cdot 10^6$
at an accelerating field of 23.4 MV/m and an average beam current of
9~mA (during the rf pulse). The  cavity has a filling time of 500
$\mu$s and a ``flat-top'' duration of 800 $\mu$s during which the
bunched beam is accelerated.  The nominal pulse repetition rate is 5~Hz. 
The time constant of the cavity equipped with a high-power coupler
is dominated by the external quality factor and practically
insensitive to the intrinsic quality factor $Q_0$ since
$$ Q_L=(1/Q_0+1/Q_{ext})^{-1}\approx Q_{ext} \quad {\rm for} \quad
Q_{ext}\ll Q_0 \;.$$ 
Therefore,  $Q_0$ cannot be derived from the time decay of the stored
energy but instead  has to be calculated from the heat transfer to the
helium bath which can be measured only with large  errors at low fields.
 
\begin{figure}[htbp]
  \centering
  \includegraphics[angle=-90,width=11cm]{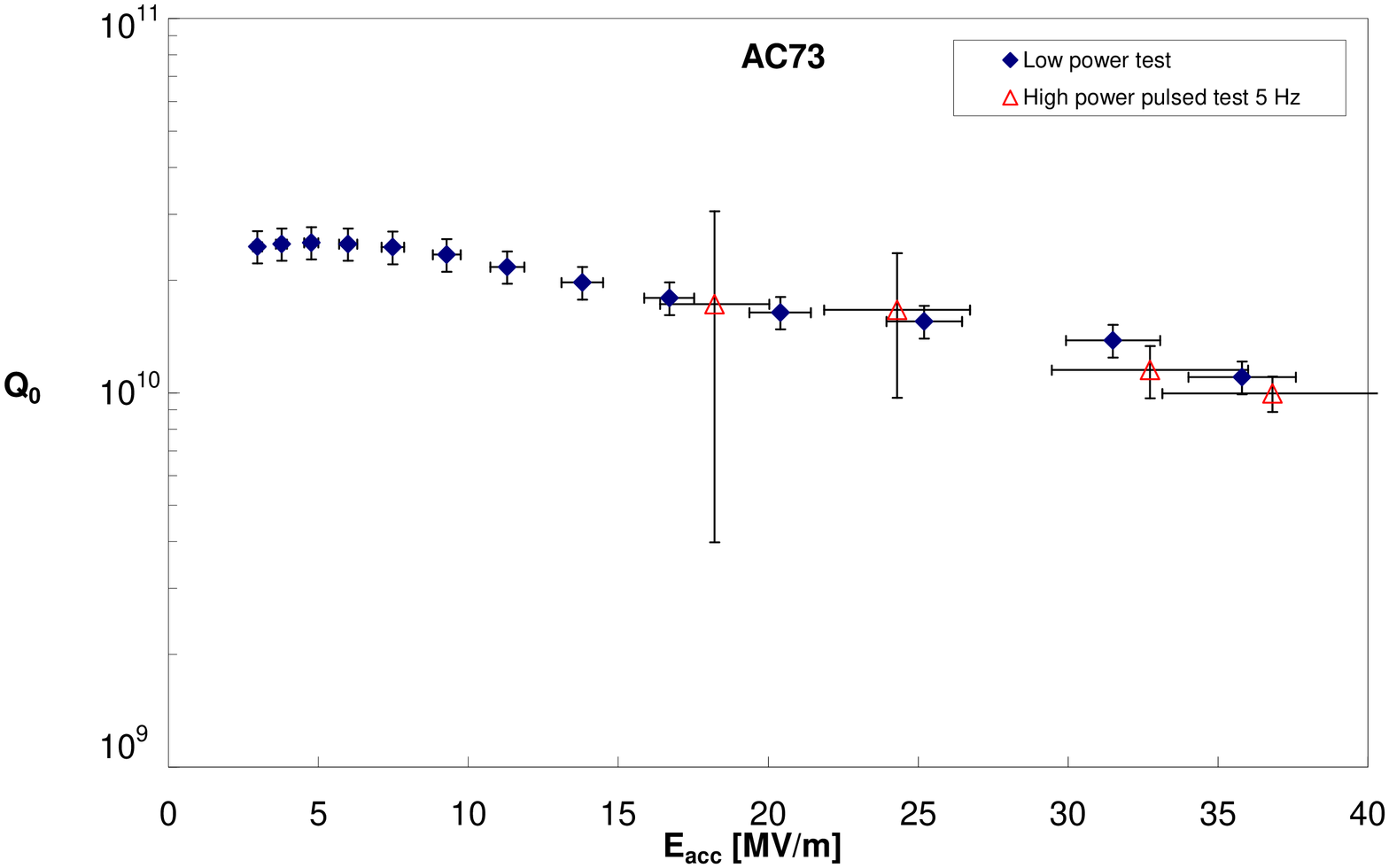}  \newline 
        a)

  \includegraphics[angle=-90,width=11cm]{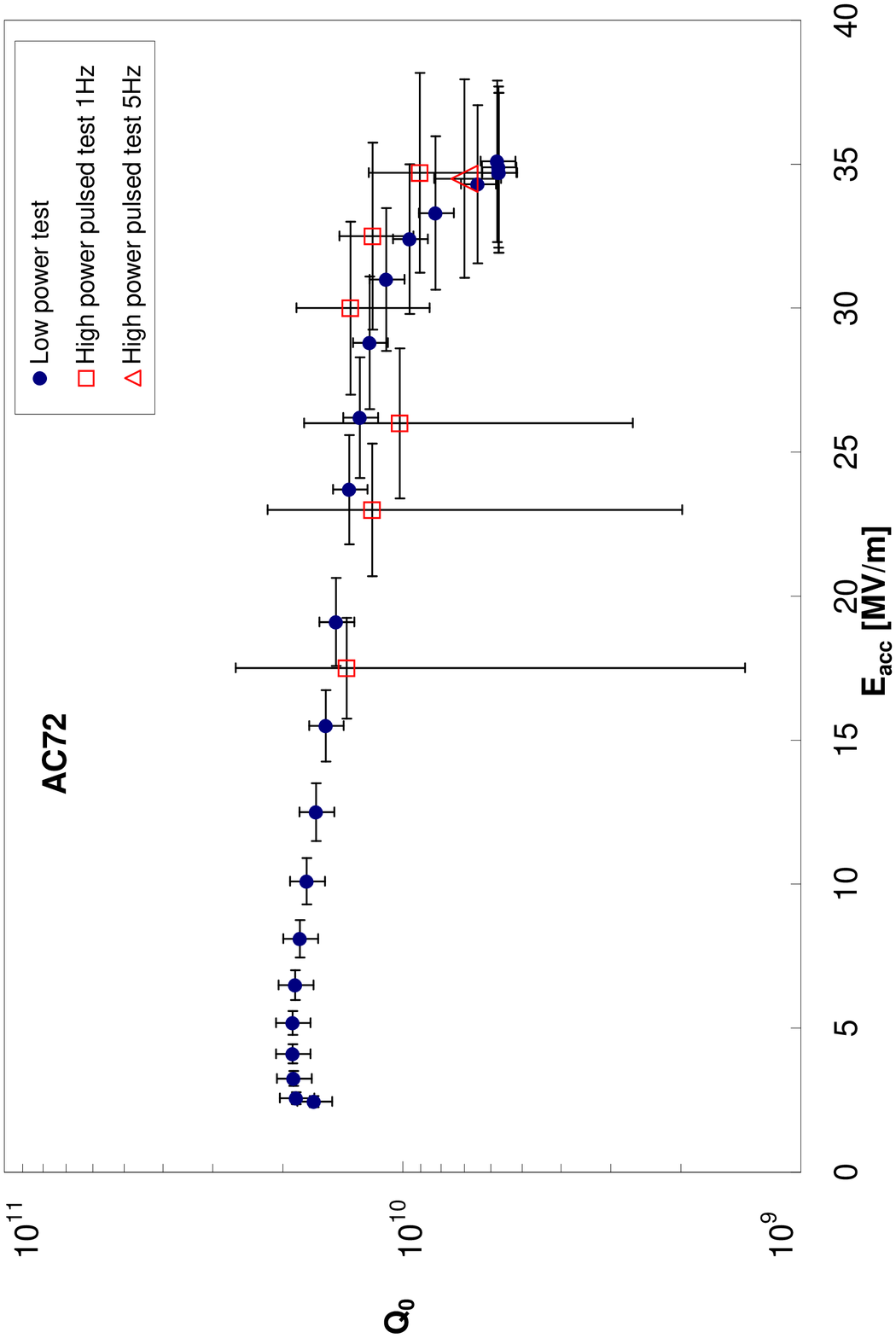}  \newline 
        b)
  \caption{High power test of two electropolished nine-cell cavities:
 (a) Cavity AC73, this cavity was operated for more than 1100
 hours at 35 MV/m. 
 (b) Cavity AC72. 
 The excitation curves obtained in the low power test in the vertical
 bath cryostat are shown for comparison and prove that the
excellent performance  is preserved after welding of
the helium tank and assembly of the high power coupler.  }
  \label{fig:ep_pulsed_test}
\end{figure}

So far, two cavities (AC72, AC73) were prepared for a high power
test without electron beam. The cavities were welded into a liquid helium tank and equipped
with a high power coupler and a frequency tuning mechanism. The tests
have been carried out in a horizontal cryostat at the TESLA Test
Facility.  Figure \ref{fig:ep_pulsed_test} shows the test results at a
repetition rate of 5 Hz for AC73 and of 1 Hz for AC72 in comparison
with the excitation curves measured in the low-power tests. It is very
encouraging that both cavities achieve the same maximum gradient as in
the low-power test. Within the large errors also the quality factors
are in agreement. Another very important result is that cavity AC73
could be operated at 35~MV/m for more than 1100 hours and
at 36~MV/m for 57 hours without any degradation.

Cavity AC73 was tested three times at a repetition rate of 10 Hz.
Between the first two tests the cavity was warmed up to room temperature,
between tests 2 and 3 it was kept for several  hours at 150 K to check
for a possible $Q$ degradation through the formation of niobium
hydrides. No indication of the ``Q-disease'' was seen. In all three
10~Hz tests the same high performance as at 5 Hz was achieved.  In
cavity AC72 a lower quality factor was seen at repetition frequencies
above 1~Hz. This could be traced back to excessive heating at a
damaged cable connector of a higher-order mode coupler.

\subsubsection{Frequency stabilization in pulsed operation}
\begin{figure}[h!]
  \centering \includegraphics[angle=0,width=10cm]{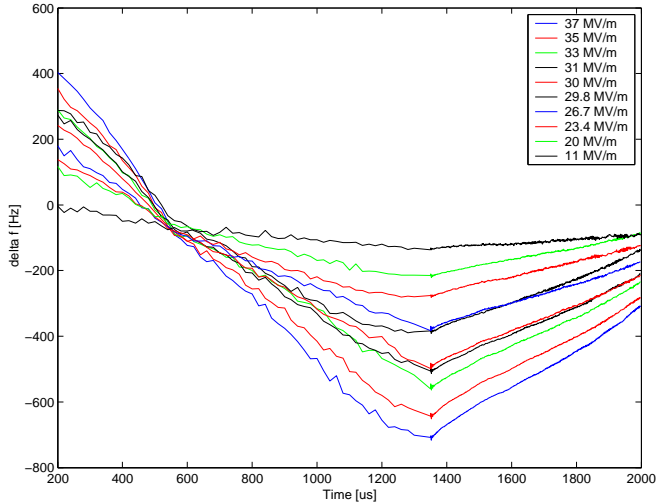}
  \caption{ Lorentz-force
  detuning in pulsed mode operation at gradients between 11 and 37
  MV/m. The resonance frequency shift $\Delta f$ is plotted as a
  function of time. Filling time of cavity between 0 and 500 $\mu$s,
  ``flat top'' between 500 and 1300 $\mu$s, decay of cavity field for
  $t>1300~\mu$s.}
  \label{fig:ep_pulsed_test_LF_detuning}
\end{figure}

The Lorentz force between the rf magnetic field and the induced
currents in a thin surface layer causes a slight deformation of the
cells in the order of micrometers and a shift in resonance
frequency which is proportional to $E_{acc}^2$.  In the pulsed
operation of the 9-cell cavities this results in a time dependent
frequency shift during the rf pulse. The TESLA cavities are reinforced
by stiffening rings which are welded between neighbouring cells and
reduce the detuning by a factor of two. Experimental data on the
detuning are shown in Fig. \ref{fig:ep_pulsed_test_LF_detuning}.  The
``Lorentz force detuning'' can be handled adequately by the rf 
system up to the nominal TESLA-500 gradient of 23.4 MV/m.

\begin{figure}[h!]
  \centering
  \includegraphics[angle=-90,width=10cm]{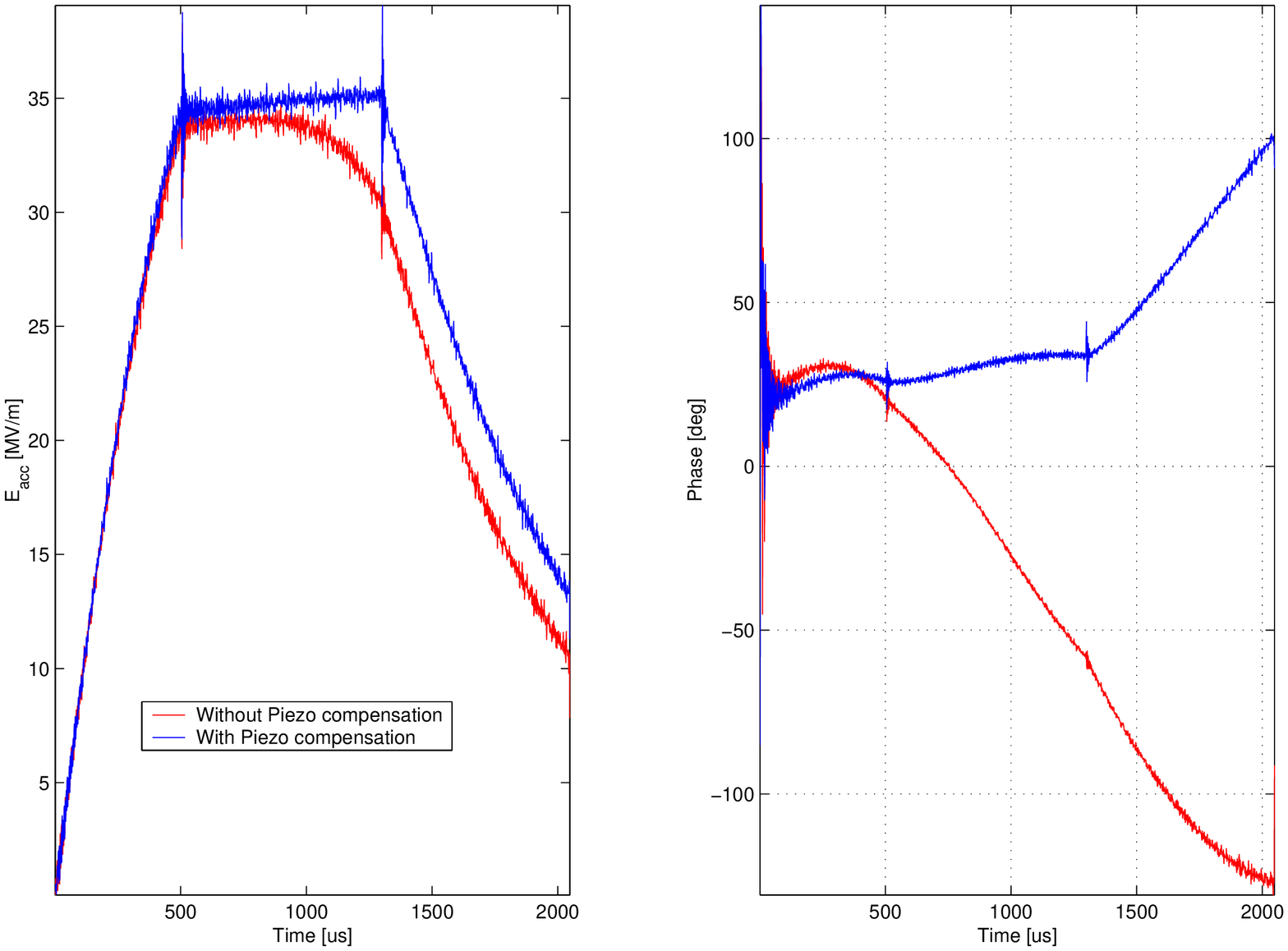}   \newline 
        a)

  \includegraphics[angle=-90,width=10cm]{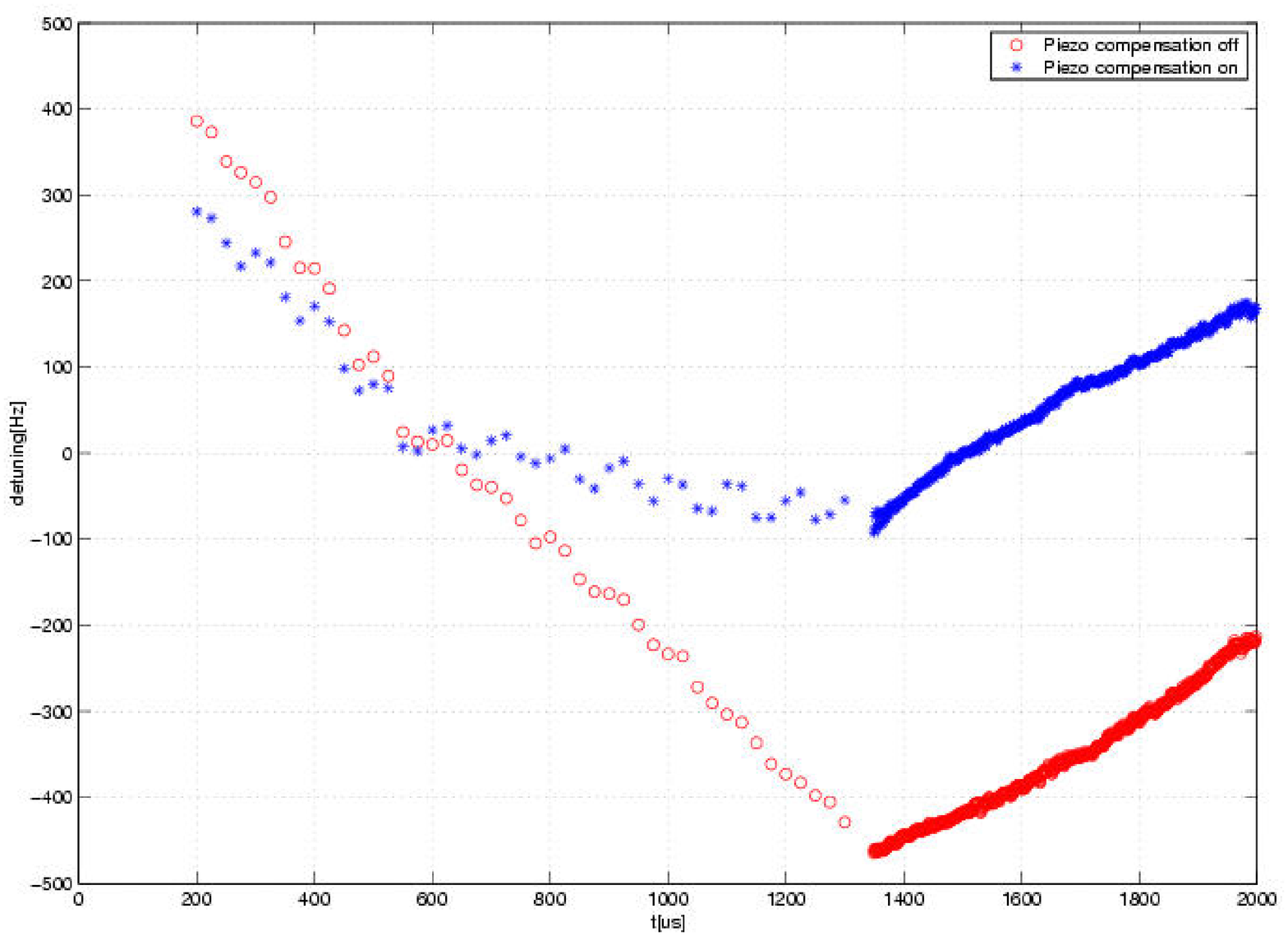}   \newline 
        b)

  \caption{High-power pulsed test of an EP cavity at 35 MV/m. 
  (a) Lorentz-force 
  detuning causes a mismatch between klystron and cavity, associated
  with a time-dependent reduction of the accelerating
  field  and a strong variation of the cavity phase
  with respect to the rf frequency. The   compensation of the
  cavity  detuning by a piezoelectric actuator leads to a constant
  accelerating field and a reduced drift in the  relative phase. 
  (b) The measured detuning at 35 MV/m 
 with and without piezo-electric    compensation.}
  \label{fig:ep_pulsed_test_LF_compensation}
\end{figure}

To allow for higher gradients the stiffening must be improved, or
alternatively, the cavity detuning  must be compensated. The latter
approach has been successfully demonstrated using a piezoelectric
tuner, see Fig.  \ref{fig:ep_pulsed_test_LF_compensation}. The
piezo-actuator changes the cavity length dynamically by a few $\mu$m
and stabilizes the resonance frequency to better than 100~Hz during the
flat-top time. 
The data in Fig. \ref{fig:ep_pulsed_test_LF_compensation} prove that
the  stiffening rings 
augmented by a piezoelectric tuning system will permit to  operate the
electropolished cavities at fixed resonance frequency up to the TESLA-800
gradient of 35~MV/m. In addition, the piezoelectric actuator may
be used to cancel  microphonic noise between the rf pulses.

\newpage

\section{Discussion of the results and conclusions}
A comprehensive understanding why electropolishing is so much superior
to chemical etching is still missing, however a few explanations
exist.  The sharp ridges at the grain boundaries of an etched niobium
surface may lead to local enhancements of the rf magnetic field and
thereby cause a premature breakdown of superconductivity at these localized
spots. A model based on this idea was developed by Knobloch et
al. \cite{liepe_99} and can account for the reduction of the quality
factor $Q_0$ at high field.  Magnetic field enhancements will be much
smaller on the smooth EP surface. Another advantage of a mirror-like
surface is that a surface barrier of the Bean-Livingston type
\cite{bl_barrier} may exist. The surface barrier prevents the
penetration of magnetic flux into the bulk niobium in a certain range
beyond the lower critical field $B_{c1}$ ($\approx 160~$mT for niobium
at 2 K).  The ``penetrating field'' $B_{pen}$ exceeds $B_{c1}$ considerably
for a perfectly smooth surface.  Magnetic fluxoids
will enter and leave the material only above this penetrating field,
hence the power dissipation associated with fluxoid motion will start
only at $B_{rf}>B_{pen}$.  The delayed flux penetration was
experimentally verified in electropolished samples of the type II 
superconductors Pb-Tl and Nb$_{0.993}$O$_{0.007}$
\cite{tomasch_64,blois_64}. The experiments showed also that
roughening of the surface by scratching and chemical etching destroyed
the barrier and reduced the penetrating field to $B_{pen} =
B_{c1}$. From these results we may conclude that an EP-treated 
superconducting cavity is likely to remain in the Meissner phase up to
an rf magnetic field exceeding $B_{c1}$ by a significant amount
whereas a BCP-treated cavity will go into the mixed phase at $B_{c1}$
and then suffer from enhanced power dissipation.

The two arguments
for the superiority of electropolished cavities refer to the
topological structure of the surface: it is smooth for EP and rough
for BCP. However, the positive influence of the 120$^\circ$C
bakeout on the high-field performance of EP-cavities cannot be
explained that way because the surface topology will not be changed 
by the bakeout.  For a discussion of
conceivable physico-chemical processes occuring during the bakeout we
refer to the review talk by B. Visentin at the recent SRF2003 workshop
\cite{visentin_03}.

In summary we can say that electropolished bulk niobium 
cavities offer the high accelerating
gradients which are required for the upgrade of the TESLA collider to
800 GeV.  For the first time, accelerating fields of 35 MV/m have been
achieved in nine-cell cavities. One cavity could be excited to 39 MV/m
even without the 1400$^{\circ}$C heat treatment, which is an
indispensible prerequisite for the BCP-treated TTF cavities to reach
25 MV/m.  In two high-power tests it
could be verified that EP-cavities preserve their excellent
performance after welding into the helium cryostat and assembly of the
high-power coupler. One cavity has been operated for 1100~hours at the
TESLA-800 gradient of 35 MV/m and 57 hours at 36 MV/m without loss 
in performance.

\section{Acknowledgments}
We thank P.~Kneisel and B.~Visentin for stimulating discussions
on the  bakeout effect.

We would like to thank  the following people for their help with the
high power rf tests and piezo operation: C. Albrecht, V. Ayvazyan,
A. Bosotti, J. Eschke, A. Goessel, R. Lange, R. Paparella,
H.-B. Peters, P. Sekalski and the DESY groups MKS3, MVP, MHF-P, MHF-SL and
MKS1.

\bibliographystyle{elsart-num}
\bibliography{LL_bibliography}

\begin{thebibliography}{10}
\expandafter\ifx\csname url\endcsname\relax
  \def\url#1{\texttt{#1}}\fi
\expandafter\ifx\csname urlprefix\endcsname\relax\def\urlprefix{URL }\fi

\bibitem{tesla_cavity_tdr}
R.~Brink\-mann, K.~Fl\"ottmann, J.~Rossbach, P.~Schm\"user, N.~Walker, H.~Weise
  (Eds.), TESLA - Technical Design Report, Vol.~II, DESY, 2001, {DESY 2001-011,
  ECFA 2001-209, TESLA Report 2001-23}.

\bibitem{tesla_cavities}
B.~Aune, et~al., {The Superconducting TESLA Cavities}, Phys. Rev. ST-AB 3~(9),
  092001.

\bibitem{kako_99}
E.~Kako, K.~Saito, et~al., {Improve\-ment} of {Cavity} {Per\-formance} in the
  {Saclay/\-Cornell/\-Desy's} sc {Cavities}, in: Krawcyk  \cite{srf_99}, pp.
  179--186.

\bibitem{lilje_99}
L.~Lilje, et~al., {Electropolishing and in-situ Baking of 1.3 GHz Niobium
  Cavities}, in: Krawcyk  \cite{srf_99}, pp. 74--76.

\bibitem{lilje_03}
L.~Lilje, et~al., {Improved Surface Treatment of the Superconducting TESLA
  Cavities}, Nucl. Inst. Meth. A 516~(2-3) (2004) 213--227.

\bibitem{gmelin_nb}
Gmelin, Handbuch der anorganischen {Chemie}, Vol. 49(Nb), Springer Verlag
  Berlin, 1970.

\bibitem{siemens_scrf}
B.~Hillenbrand, N.~Krause, K.~Schmitzke, Y.~Uzel, {Abschlussbericht -
  Supraleitende Resonatoren}, Tech. Rep. NT 2024 7, Siemens AG, {BMBF
  Forschungsbericht} (Dezember 1982).

\bibitem{guerin_bcp2}
J.~Guerin, Etude du bain de polissage chimique de niobium, Tech. Rep.
  TE/LC/157/89, CERN (October 1989).

\bibitem{kneisel_kfk}
P.~Kneisel, Surface preparations of niobium, in: M.~Kuntze (Ed.), Proceedings
  of the Workshop on RF Superconductivity, Vol. I+II, KFK, Karlsruhe, 1980,
  p.~27, {KFK-3019}.

\bibitem{ponto}
L.~Ponto, M.~Hein, Elektropolitur von {Niob}, Tech. Rep. {WUP 86-17}, Bergische
  Universit\"at Wuppertal (1986).

\bibitem{claire_99}
C.~Z. Antoine, et~al., Alternative approaches for surface treatment of {Nb}
  superconducting cavities, in: Krawcyk  \cite{srf_99}, pp. 109--117.

\bibitem{saito_ep_system}
K.~Saito, et~al., Electropolishing of {L}-band cavities, in: Y.~Kojima (Ed.),
  Proceedings of the 4th Workshop on RF Superconductivity, Vol. I+II, KEK,
  Tsukuba, 1989, pp. 635 -- 695, {KEK Report 89-21}.

\bibitem{visentin_bake_98}
B.~Visentin, J.~Charrier, B.~Coadou, Improvements of superconducting cavity
  performances at high gradients, in: Proceedings of the 6th EPAC, Vol. III,
  Stockholm, 1998, p. 1885.

\bibitem{kneisel_99}
P.~Kneisel, Preliminary experience with ''in-situ'' baking of niobium cavities,
  in: Krawcyk  \cite{srf_99}, pp. 328--335.

\bibitem{saito_99}
K.~Saito, et~al., {High Gradient Performance by Electropolishing with 1300 MHz
  Single and Multi-cell Niobium Superconducting Cavities}, in: Krawcyk
  \cite{srf_99}, pp. 288--291.

\bibitem{liepe_99}
J.~Knobloch, M.~Liepe, R.~Geng, H.~Padamsee, {High-Field Q slope in
  Superconducting Cavities Due to Magnetic Field Enhancement at Grain
  Boundaries}, in: Krawcyk  \cite{srf_99}, pp. 77--91.

\bibitem{bl_barrier}
C.~Bean, J.~D. Livingston, {Surface Barrier in Type-II Superconductors}, Phys.
  Rev. Lett. 12~(1) (1964) 14--16.

\bibitem{tomasch_64}
A.~Joseph, W.~J. Tomasch, {Experimental Evidence for Delayed Entry of Flux into
  a Type-II Superconductor}, Phys. Rev. Lett 12~(9) (1964) 219--222.

\bibitem{blois_64}
R.~D. Blois, W.~de~Sorbo, {Surface Barrier in Type-II Superconductors}, Phys.
  Rev. Lett. 12~(18) (1964) 499--501.

\bibitem{visentin_03}
B.~Visentin, {Q-slope at high gradients: Review about experiments and explanations}, in: Proceedings of 11th Workshop Superconducting RF, Luebeck, 2003.

\bibitem{srf_99}
F.~Krawcyk (Ed.), Proceedings of the 9th Workshop on RF Superconductivity, Vol.
  I+II, LANL, Santa Fe, 1999.

\end{thebibliography}

\end{document}